\documentclass[aps,superscriptaddress,twocolumn,showpacs]{revtex4}
\usepackage{bm,graphicx,amsmath,amssymb,bm}
\input{tcilatex}

\begin{document}

\title{Slow relaxation of low temperature vortex phases in Bi$_{2}$Sr$_{2}$CaCu$_{2}
$O$_{8}$}
\author{F. Portier}
\affiliation{Service de Physique de l'Etat Condens{\'e}, 
Commissariat \`a l'Energie Atomique, Saclay, F-91191 Gif-Sur-Yvette, France}
\author{G. Kriza}
\affiliation{Research Institute of Solid State Physics, POB.49 H-1525 Budapest, Hungary}
\author{B.Sas}
\affiliation{Research Institute of Solid State Physics, POB.49 H-1525 Budapest, Hungary}
\author{L.F. Kiss}
\affiliation{Research Institute of Solid State Physics, POB.49 H-1525 Budapest, Hungary}
\author{I. Pethes}
\affiliation{Research Institute of Solid State Physics, POB.49 H-1525 Budapest, Hungary}
\author{K. Vad}
\affiliation{Institute of Nuclear Research, POB.51 H-4001 Debrecen, Hungary}
\author{B. Keszei}
\affiliation{Research Institute for Materials Science, POB.49 H-1525 Budapest, Hungary}
\author{F.I.B.Williams}
\affiliation{Service de Physique de l'Etat Condens{\'e}, 
Commissariat \`a l'Energie Atomique, Saclay, F-91191 Gif-Sur-Yvette, France}

\date{\today}

\begin{abstract}
Nonlinear transport in the low temperature vortex glass state of single
crystal Bi$_{2}$Sr$_{2}$CaCu$_{2}$O$_{8}$ has been investigated over long
times (up to two weeks) with fast current pulses driven along the $ab$
plane. It is found that at low temperature and high magnetic field both zero
field cooled (ZFC) and field cooled (FC) samples relax towards the same
equilibrium state, which is much closer to the original ZFC state than to
the FC state. The implication of this FC metastability for the vortex phase
diagram is discussed in the context of previous measurements.
\end{abstract}

\maketitle

The interest in vortices in strongly anisotropic layered high $T_{c}$
superconductors like BSCCO (Bi$_{2}$Sr$_{2}$CaCu$_{2}$O$_{8}$) arises
firstly from the weak interplane coupling which makes the basic vortex
entity a quasi-2-D vortex segment (``pancake vortex'') within a layer and
renders the interaction-disorder problem particularly rich by the interplay
between intraplane repulsion, interplane attraction, random bulk pinning and
thermal fluctuations \cite{lawrence,koshelev}. Secondly the configuration and
dynamics of the vortices play a vital role in the current carrying capacity
of the superconductor and in this respect the low temperature pinned vortex
solid phases are expected to be the most resistant to forces from transport
currents.

Recently an unexpected phenomenon of metastability and conversion has been
reported \cite{BSCCO1} for this phase in BSCCO, recalling what was found in
two low $T_{c}$ superconductors, NbSe$_{2}$ \cite{henderson} and pure Nb \cite
{ling} where the field cooled (FC) preparation (sample cooled from above $%
T_{c}$ with field applied) is metastable below a certain temperature and can
be converted into a state resembling that produced by a zero field cooled
(ZFC) preparation (sample cooled in zero field and field applied only on
reaching the measurement temperature). Motivated by the possible coexistence
of competing low temperature solid phases and by aging of a semi-ordered
system in a random field we have followed the evolution over long times of
both FC and ZFC prepared states in BSCCO .

The experiment probes the state of the vortex system by measuring the
voltage response to short - to avoid heating - high current (25-100$\mu$s,
maximum $\sim $1.3$\times $critical current) triangular pulses. The
threshold current for dissipation is taken as the principal indicator of the
state of the vortices. We distinguish, conceptually, non-equilibrium
stationary states, such as the Bean critical state, from thermodynamically
stable and metastable equilibrium states by the time scales: measurements of
the threshold current are made in $\sim 10^{-5}$s whereas stationary Bean
type density profiles from vortex transport decay typically in $
10^{2}-10^{3}$s while the relaxation discussed in this paper occurs in $%
10^{3}-10^{6}$s. For the present purposes, we define a metastable state as
showing equilibrium over $10^{3}-10^{4}$s.

\begin{figure}[tbp]
\includegraphics*[width=6 cm]{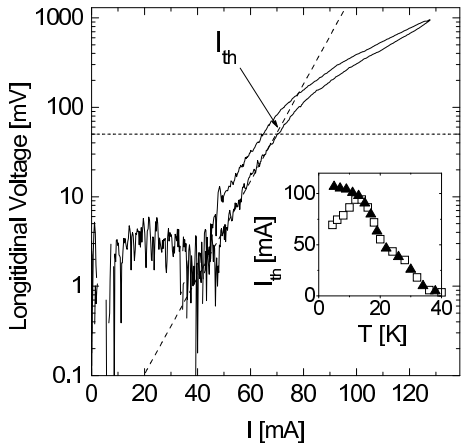}
\caption{$VI$ characteristic for a Zero Field Cooled sample at $T=4.5$K, $%
\protect\mu _{0}H=1.5$T at $t=0$. The long-dashed line is a Kim-Anderson fit
to the data. The dotted line indicates the definition of $I_{th}$ used in
this study. The inset shows the temperature variations of $I_{th}^{FC}$ $%
(\blacktriangle )$ and $I_{th}^{ZFC}$ $(\square )$ for $\mu_{0}H=1.5$T for freshly prepared samples.}
\label{Ilogv}
\end{figure}

Previously \cite{BSCCO1}, we reported $ab$ plane measurements of the $VI$
characteristics of freshly FC and ZFC prepared BSCCO monocrystals at large
currents, using the same fast current pulse technique. For perpendicular
magnetic fields $H > 1$kOe and low temperatures $T<12$K, the $VI$
characteristic showed FC to result in a higher threshold current $I_{th}$
than ZFC. On warming, $I_{th}^{ZFC}$ increases and $I_{th}^{FC}$ decreases
to merge at a temperature $T_{p}(H)$ beyond which the $VI$ characteristic no
longer depends on preparation (insert of Fig. \ref{Ilogv}). Unlike ZFC, the
FC threshold current is not a unique function of $(H,T)$ but depends on the
cooling rate and is found to be metastable in the sense that a small
temporary perturbation of the field $\Delta H\simeq \pm 300$Oe induces a
change in the $VI$ characteristic which though it depends on the duration of
the perturbation always tends towards the ZFC characteristic. Furthermore if
one ZFC prepares a sample at low temperature and warms, the threshold is the
same as for direct ZFC preparation at that warmer temperature; however on
subsequent cooling it does not return along the ZFC line, but instead
behaves rather like the FC line, with lower starting value and similar
metastability.

We present data for two melt cooling fabricated single crystals \cite{Keszei}
of about $1\times 0.5\times 0.00$3mm$^{3}$ oriented with $\ c-$axis parallel
to the magnetic field and critical temperatures of 83K and 81K with a
transition width of $\sim 2$K at zero field. The anisotropy coefficients $%
\gamma \approx 500$ were estimated from normal state resistivity at $90$K
where $\rho _{ab} \approx 100\mu \Omega $cm . The current pulses were
injected and withdrawn through two symmetrically placed contacts in a way to
keep the mean potential of the sample at ground; the potential drop was
measured between two symmetrically interspaced potential contacts with a
differential amplifier . The temperature was electronically regulated and
the magnetic field provided by a superconducting magnet in persistant mode.
FC samples were cooled from above $T_{c}$ at 240K h$^{-1}$
to 30K and at 60K h$^{-1}$ thereafter. ZFC samples
were cooled in the same way before applying the field at 100 Oe s$^{-1}$ .

\begin{figure}[tbp]
\includegraphics[height=8.6cm,angle=270]{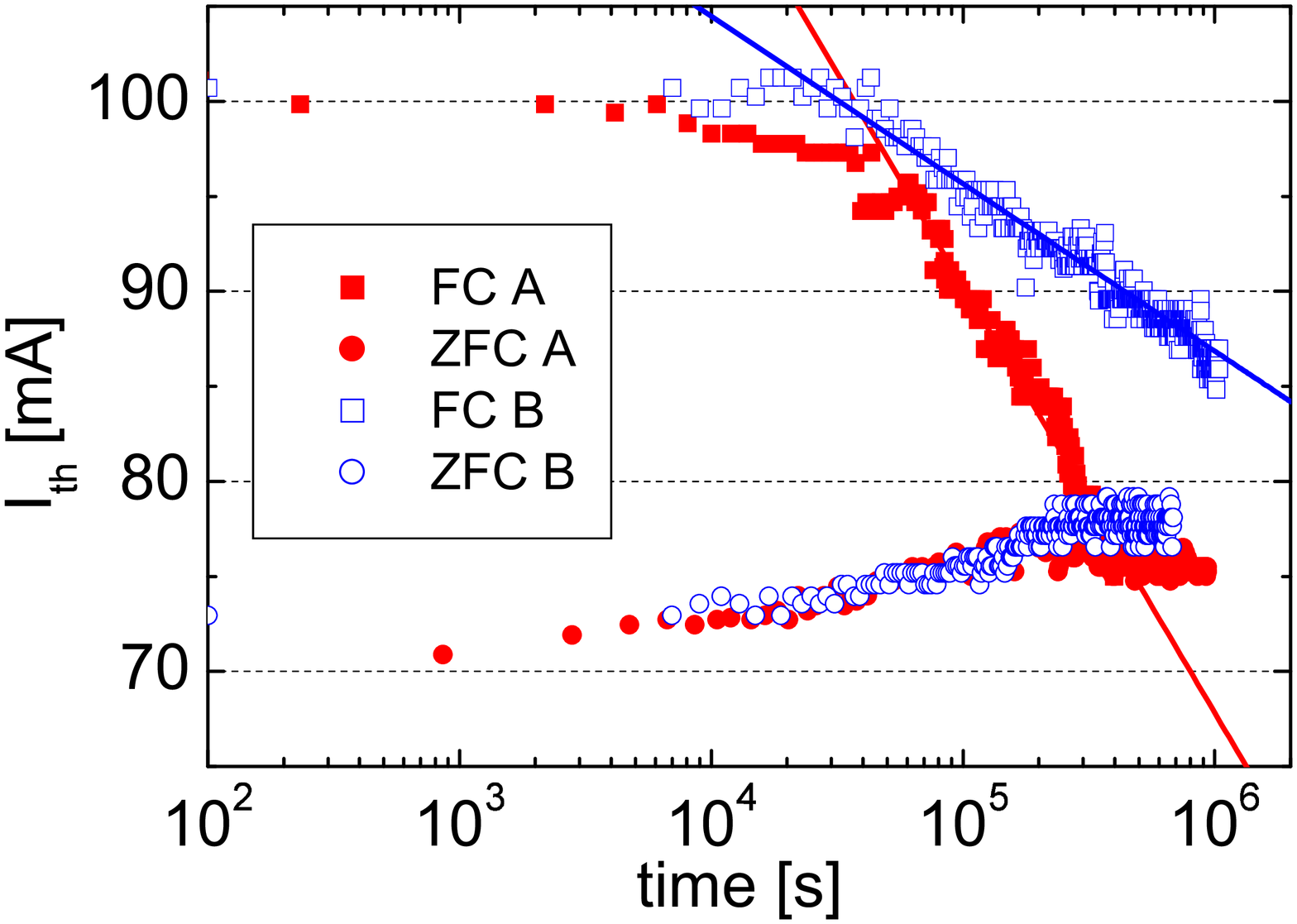}
\caption{Evolution of $I_{th}$ at $T=4.5$K and $\protect\mu _{0}H=1.5$T on
two different samples sample A corresponds to the red points and sample B to
the blue points. The threshold current is interrogated with short 
$25 \mu$s or $100 \mu$s
triangular pulses every $10^{3}$s.}
\label{Ilogt}
\end{figure}

The first experiments (Fig.\ref{Ilogt}) showed the $VI$ characteristics of
FC and ZFC samples to relax towards one another . To characterise the state
of the sample we choose the threshold current $I_{th}$ which signals when
the force exerted by the transport current depins the vortices in the top
superconducting plane. But because the shapes of FC and ZFC $VI$ are
different \cite{BSCCO1}, a modification of the form accompanies the
relaxation and requires us to define $I_{th}$ to have the same meaning
throughout the relaxation. Since from the $VI$ plot in Fig.\ref{Ilogv} we
see that at voltages $V<V_{th}\simeq 50\mu$V both FC and ZFC $VI$ are well
described by a Kim-Anderson type relation \cite{anderson}, we define the
threshold current by $V(I_{th})=50$ $\mu$V. Although the exact
voltage value is somewhat arbitrary, the dependence of $I_{th}$ on $V_{th}$
is only logarithmic. The inset of Fig. \ref{Ilogv} shows the temperature
variation of $I_{th}^{FC}$ and $I_{th}^{ZFC}$ at $t=0$ for $\mu _{0}H=1.5$T.

Fig.\ref{Ilogt} shows the time evolution of $I_{th}$ for FC and ZFC
preparations at $T=4.5$K and $\mu _{0}H=1.5$T for two different samples with
different threshold currents. One sees immediately that the total variation
of $I_{th}^{FC}$ is much greater than that of $I_{th}^{ZFC}$ and that the
aging produces the same ultimate effect as a short term field perturbation 
\cite{BSCCO1} including the diminishing of the $IV$ hysteresis. The results
for the second sample (blue curve) have been scaled by a multiplicative
factor of $0.45$ which maps the intial evolution onto that of the first
sample. Both show a delayed logarithmic decay in\textbf{\ }$%
I_{th}^{FC}=I_{0}-I_{1}\log _{10}{t[}$s${]}$ after remaining nearly constant
for $\sim 10^{4}$s. For the first sample $I_{th}^{FC}(t)$ evolves until it
meets $I_{th}^{ZFC}(t),$after which it becomes indistinguishable from it.
The second sample behaves similarly except that the experiment was curtailed
before the two curves met. The ZFC threshold varies only weakly and on the
same $10^{2}-10^{3}$s time scale as in the Josephson Plasma Resonance 
\cite{matsuda} and AC Campbell length \cite{prozorov} experiments. On this time
scale, which is compatible with relaxation of the Bean profile seen in the
direct observation of local induction \cite{berry}, there is no variation of
a FC sample.

\begin{figure}[tbp]
\includegraphics[height=8.6cm,angle=270]{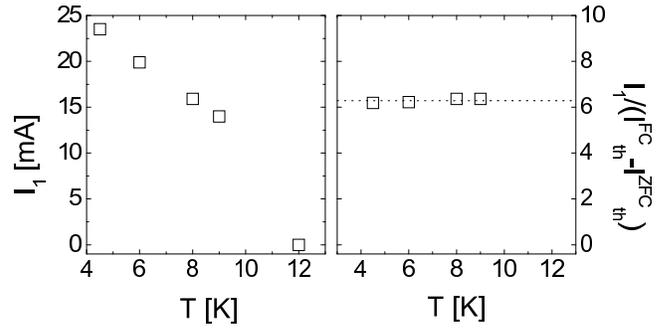}
\caption{Left: relaxation rate $I_{1}=|dI_{th}^{FC}/d\log _{10}(t)|$. Right: 
$I_{1}$ normalized by $(I_{th}^{FC}(t=0)-I_{th}^{ZFC}(t=0))$. These data
were collected on sample A, using the current pulse sequence leading to the
red points in fig. \ref{Ilogt}. For $T\geq 12K$, no relaxation is observed
and $I_{th}^{FC}=I_{th}^{ZFC}.$}
\label{I1}
\end{figure}

What drives the conversion? We note first that the Bean-Livingston surface
barrier \cite{bean-livingston}, which can block the flow of vortices,
vanishes above $H_{c}\simeq 10^{3}$Oe \cite{fuchs}, whereas we have observed
FC metastability and relaxation up to $H=1.8\ 10^{5}$Oe. The relaxation is
slower the higher the temperature: the rate $I_{1}=|dI_{th}^{FC}/d\log
_{10}(t)|$ shown on Fig.\ref{I1}, \textbf{decreases} linearly with \textbf{%
increasing} temperature and \textbf{vanishes }at $T_{p}$, the temperature at
which $I_{th}^{ZFC}(t=0)$ reaches its maximum and where the metastability
evoked earlier \cite{BSCCO1} ceases. Remarkably the ratio $%
I_{1}/(I_{th}^{FC}(t=0)-I_{th}^{ZFC}(t=0))$ remains constant within $5\%.$
The interrogation pulses themselves seem to have no effect: pulses giving a
vortex displacement estimated from $s=\int v_{y}dt=c\int E_{x}/B_{z}dt\geq
(c/B)\int V(t)dt$ greater than the width of the sample do not modify $%
I_{th}$ for either preparation, indicating that the remnants of a critical
density profile left by the drive current do not alter the difference
between the two states. We checked that the interrogation pulse sequence
does not induce aging by verifying that a single or $2.10^{3}$ alternated
phase full wave 25$\mu$s, or $2.10^{3}$ unipolar half wave 100$\mu$s 
pulses per day gave the same result on the same sample. 
Each 100$\mu$s half-wave pulse induces an average vortex 
displacement of $s \sim 10 \mu$m while the 25$\mu$s
full wave pulse displaces a vortex by $\sim 10$nm. Alternating
the phase eliminates directional bias so that the total displacement should
be a random sum. Although the blue and red symbols on Fig. \ref{Ilogt} were
obtained with different samples, the form of the evolution is the same; when
all the values are rescaled by a single constant, the ZFC evolution is
identical for both samples whilst the FC starts identically with about the
same delay before decaying with about half the logarithmic slope. It was
also established that field variation catalyses aging: in experiments in a
very stable magnet ($1.10^{-4}$ day$^{-1}$ decay) reset every day, no
evolution was detectable in any of the three interrogation procedures
whereas application of a small field step makes the FC threshold age \textbf{%
monotonically towards the ZFC value }whereas if the variation in threshold
were to result from a Bean-type density profile, it would be expected to
return towards its FC value as the profile diffuses away. Fig. \ref{Ilogt}
should be viewed as a convolution of the basic response to a field step with
the natural decrease in magnetic field of the magnet used ($2.10^{-3}$ day$%
^{-1}$). Bearing in mind that, due to the anisotropy, tipping the sample by $%
10^{-3}$ rad gives rise to screening current densities of $\sim 10^{5}$ A cm$%
^{-2},$ we were careful about mechanical stability; indeed a careless
cryogen transfer sometimes reduced $I_{th}^{FC}$ for several minutes, but
unlike mechanical shock on NbSe$_{2}$ \cite{henderson}, did not result in
permanent conversion.

Without further reasoning, we conclude that the FC state is metastable at
low temperatures and decays to a state which, judging by the threshold
current, is very similar to the ZFC state, that the conversion is catalysed
by a small field variation and is slower the smaller the difference in
threshold currents and that Bean type density profiles do not play a
significant role.

\begin{figure}[tbp]
\includegraphics[height=7cm,angle=270]{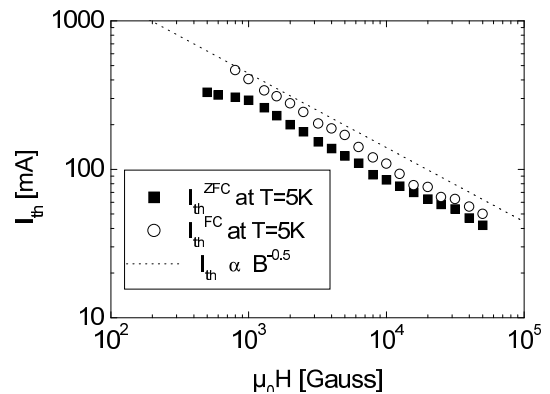}
\caption{FC and ZFC threshold current $I_{th}$ at $T=5$K as a function of $%
\protect\mu _{0}HM.$ The dotted line shows $H^{-0.5}.$ }
\label{IB}
\end{figure}

To say more we must understand the meaning of the lower threshold current
for ZFC and this in turn requires an evaluation of the relative
contributions of $c$-axis and $ab$-plane currents. We have shown in other
experiments \cite{current} that the resistive front spreads out from the
contacts, indicating that the current penetrates progressively into the
layers. This implies that phase slips occur between adjacent superconducting
planes and the $VI$ characteristic reflects a mixture of $c$-axis and $ab$%
-plane transport properties \cite{current,khaykovich3}. Thus although
the voltage is a direct measure of the vortex velocity in the top
superconducting plane, the threshold current reflects \textbf{both }$ab$%
-plane pinning \textbf{and} $c$-axis Josephson link properties. We can
estimate their relative importance from the magnetic field dependence of $%
I_{th}$ below $T_{p}$ shown on Fig.\ref{IB}. Both FC and ZFC samples exhibit
a power-law behavior $I_{th}\varpropto B^{-\nu }$, with $\nu ^{FC}=0.55\pm
0.01$ and $\nu ^{ZFC}=0.50\pm 0.01$. This is similar to BSCCO (2212) thin
films \cite{labdi} and BSCCO(2212)/BSC0(2201) superlattices \cite{raffy}
where, due to the small number of superconducting planes, the current
distribution is expected to be uniform ($J_{c}=0)$ and dissipation to arise
only from the motion of flux lines. On the other hand the $c$-axis critical
current \cite{suzuki} of BSCCO mesa structures of small lateral extension
showed $J_{th}^{c}\propto 1/H$ at $T=5$K and $H>10$kOe. Compatibility
imposes that the contribution to the critical current that we measure be
dominated by $J_{th}^{ab}.$ This allows us to say something about the nature
of the pinning and the dimensionality of the order. The low exponents in the
magnetic field dependence of $I_{th}$ are in disaccord with weak pinning
Larkin-Ovchinikov theory which leads to $J_{th}^{ab}\propto B^{-1}$ for both
2D and 3D vortex configurations \cite{koshelev}. Strong pinning of 2-D
ordered vortex segments where the pinned vortices remain trapped, however,
does lead to the $I_{th}\propto B^{-0.5}$ that we observe. Also, on the
grounds that greater order means less pinning, the fact that $%
I_{th}^{FC}>I_{th}^{ZFC}$ indicates that the ZFC preparation is more
ordered, as in Nb and 2H-NbSe$_{2}$ . Neutron diffraction on FC prepared
BSCCO shows low temperature ordered and disordered states respectively below
and above $\sim 500$Oe \cite{cubitt}; unfortunately no neutron measurements
on ZFC samples were reported.

Recent numerical simulations by Exartier and Cugliandolo \cite{raphael} show
the $VI$ of a disordered vortex state to have higher threshold current and
stronger hysteresis than an ordered one, consistent both with the above
argument that FC is more disordered than ZFC and with the observed
decreasing hysteresis for FC with aging.

Similar FC metastability has been observed in 2H-NbSe$_{2}$ \cite{henderson}%
, and in Nb \cite{ling}, where in addition it was shown by neutron
diffraction that the stable ZFC state is more ordered than the metastable FC
state. However there are also differences: in 2H-NbSe$_{2}$, the ZFC $I_{th}$
increases with increasing temperature only in a narrow temperature range
close to the melting line, in agreement with the weak pinning
Larkin-Ovchinikov picture of the peak effect \cite{higgins}, whereas in
BSCCO $T_{p}(H)$ $\lessapprox T_{m}/2$ \ and there is no clear sign of a
peak-effect near the melting line and the low temperature linear temperature
dependence of $I_{th}^{ZFC}$ persists to at least 80mK \cite{BSCCO1}.

A microscopic model which collates these observations starts with the idea
that the thermodynamically stable zero temperature, zero disorder ground
state must consist of an Abrikosov lattice of continuous vortex lines.
Introducing the vortices at zero temperature can result in a state close to
this in the limit of small fractional volume of pins. Upon heating, thermal
fluctuations induce the lines to explore the space about their (also
metastable) equilibrium positions and if a line encounters a pin it sticks
to it, resulting in a greater number of anchor points and greater threshold
current. However if the pinning potential is deep, the lines remain pinned
on subsequent recooling with the same sort of metastability as an FC
prepared state, giving rise to the type of hysteresis in temperature
observed for ZFC. A magnetic field variation, in changing the vortex
density, tends to push the lines out of their traps and liberate them to
reconnect as for the ZFC preparation. The peak in ZFC threshold current
corresponds either to full exploration of the disorder potential - but that
can only occur near melting - or to a new ground state. Evidence for two
distinct phases is to be found in the fact that the conversion rate is
slower nearer the peak; a single phase description would give faster
relaxation the higher the temperature whereas to access a new phase it is
necessary to find a collective minimum in the thermodynamic potential,
involving a priori unlikely simultaneous displacements of individual
vortices and, in the case of a first order transition, a potential barrier.
In this picture, an FC prepared state above $T_{p}$ occupies a maximum of
pins consistent with 2-dimensional ordering, a state which cannot find a
sufficiently probable unassisted path to the low temperature phase with
finite $c-$axis correlation.

In summary the high field vortex configuration in FC prepared BSCCO is
metastable and ages towards the ZFC preparation. This behavior, although
catalysed by field variation, is neither associated with a stationary
critical density profile nor with the measurement perturbation and because
it is slower closer to the peak in the ZFC threshold it reinforces the idea
that the metastability line of the previous study \cite{BSCCO1} reflects a \
first order phase boundary, probably connected with the absence of the
second magnetization peak \cite{khaykovich} for $T\leq 14$K \cite{vanderbeek}%
. More tentatively it appears that the ZFC preparation results in a more
ordered state and that a strong pinning scenario applies.

It is a pleasure to acknowledge experimental help from P.Jacques, technical
aid from R.Tourbot and C.Chaleil and much discussion and counsel from L.
Cugliandolo, R. Exartier, S. Bhattacharya, E. Andrei, L. Forr\'{o}, H.
Raffy, A. Pomar and E. Vincent. B.Sas and G.Kriza thank the Commissariat 
\`{a} l'Energie Atomique for a Bursary to visit the Saclay laboratory.
Research in Hungary has been supported by the grant OTKA-T029877. We are
grateful also for the support of the ``BALATON''  programme of the French
and Hungarian Foreign Affairs Ministries.

\end{document}